\documentclass{PoS}
\usepackage{amsfonts,amssymb,amsmath,bm}
\usepackage{graphicx}

\newcommand{\be}{\begin{equation}}
\newcommand{\bea}{\begin{eqnarray}}
\newcommand{\ee}{\end{equation}}
\newcommand{\eea}{\end{eqnarray}}
\newcommand{\bpi}{\begin{picture}}
\newcommand{\bce}{\begin{center}}
\newcommand{\epi}{\end{picture}}
\newcommand{\ece}{\end{center}}

\newcommand{\D}{\displaystyle}
\def\chic#1{{\scriptscriptstyle #1}}
\def\gb{{\Gamma}}
\def\g{\widetilde{\gb}}

\title{Gluon masses without seagull divergences}

\ShortTitle{Gluon masses without seagull divergences}

\author{{Joannis Papavassiliou}\\
Department of Theoretical Physics and IFIC, \\
University of Valencia-CSIC, \\
E-46100, Valencia, Spain.\\
        E-mail: \email{Joannis.Papavassiliou@uv.es}}

\abstract{

The  study  of  dynamical  gluon  mass  generation  at  the  level  of
Schwinger-Dyson equation involves a delicate interplay between various
field-theoretic  mechanisms  The  underlying  local  gauge  invariance
remains  intact by  resorting  to the  well-known Schwinger  mechanism, 
which  is assumed to  be realized 
by  longitudinally coupled bound  state poles,
produced by  the non-perturbative dynamics of the  theory. These poles
are subsequently included  into the Schwinger-Dyson equation of 
the gluon propagator through the  three-gluon vertex, generating a
non-vanishing  gluon mass, which,  however, is  expressed in  terms of
divergent  seagull   integrals.   In  this talk  we explain how  such
divergences can be eliminated completely by virtue of a characteristic
identity, valid in dimensional regularization.  The ability to trigger
this identity depends, in turn,  on the details of 
the three-gluon vertex employed, and in particular, on the exact way 
the bound  state poles are incorporated.
A concrete example of a vertex that triggers the 
aforementioned identity
is constructed, the ensuing cancellation of  all seagull divergences 
is explicitly demonstrated, and a  
finite gluon mass is obtained. 
Due to the multitude of conditions that 
must be simultaneously satisfied, this 
construction appears to be exclusively realized within the PT-BFM framework.   
The resulting  system of integral  equations gives rise to   
a gluon  mass that displays power-law  running and an 
effective charge which, due to the presence of the 
gluon mass,  freezes  in  the  infrared at a finite
(non-vanishing) value.}

\FullConference{International Workshop on QCD Green's Functions, Confinement, and Phenomenology - QCD-TNT09\\
		 September 07 - 11 2009\\
		 ECT Trento, Italy}

\begin{document}

\section{\label{mgsd} Introduction}

The  gluon is
massless  at the  level  of the  fundamental  QCD Lagrangian, and  remains
massless to all order in perturbation theory. 
However, as Cornwall argued in the early eighties~\cite{Cornwall:1982zr}, the 
non-perturbative QCD
dynamics generate  an  effective,  momentum-dependent  mass for the gluons, 
without affecting    the   local    $SU(3)_c$   invariance,    which   remains
intact.

Given that the gluon mass generation is a purely non-perturbative effect, the natural 
framework to study it, in the continuum,  are the 
Schwinger-Dyson equations (SDEs) of the theory. 
At the level of the SDEs the  generation of such a  mass is associated with 
the existence of infrared finite solutions for the gluon propagator~\cite{Cornwall:1982zr,Aguilar:2006gr,Aguilar:2008xm}.
In covariant gauges, the gluon propagator, $\Delta_{\mu\nu}(q)$, has the form  
\be 
\Delta_{\mu\nu}(q)= -i\left[ {\rm P}_{\mu\nu}(q)\Delta(q^2) +\xi\frac{\D q_\mu
q_\nu}{\D q^4}\right] \,,
\label{fprop}
\ee
where $\xi$ denotes the gauge-fixing parameter, and 
${\rm P}_{\mu\nu}(q)= g_{\mu\nu} - q_{\mu} q_{\nu}/{q^2}$.
The scalar factor $\Delta(q^2)$ is given by $\Delta^{-1}(q^2) = q^2 + i \Pi(q^2)$, 
where  $\Pi_{\mu\nu}(q)={\rm P}_{\mu\nu}(q) \,\Pi(q^2)$ is the gluon self-energy;
the dimensionless vacuum polarization, ${\bm \Pi}(q^2)$, is defined  
as $\Pi(q^2)=q^2 {\bm \Pi}(q^2)$. 
So, in general, one looks for solutions 
with  $\Delta^{-1}(0) > 0$.  
Such solutions may  be  fitted  by     ``massive''  propagators  of   the form 
$\Delta^{-1}(q^2)  =  q^2  +  m^2(q^2)$;
$m^2(q^2)$ is  not ``hard'', but depends non-trivially  on the momentum  transfer $q^2$.

The pinch technique (PT) propagator, 
usually denoted by $\widehat\Delta(q^2)$ in the literature~\cite{Cornwall:1982zr,Cornwall:1989gv}, 
is the ideal quantity to study in this context, because 
it is independent of the gauge-fixing parameter ($\xi$). Therefore, 
any statement about its infrared behavior, and in particular 
the generation of a gluon mass, is bound to be free of gauge artefacts.
In recent studies, however, the tendency has been 
to focus on the gluon propagator in a fixed gauge, such as the 
conventional Landau gauge ($\xi=0$), instead of the privileged  
Feynman gauge ($\xi_Q=1$) of the background field method (BFM)~\cite{Abbott:1980hw}, 
which, quite remarkably,  
reproduces automatically the PT results (third item in~\cite{Cornwall:1989gv}). 
The main reason for this less optimal choice is 
the need to compare meaningfully the SDE results with those obtained from 
lattice studies, which, almost exclusively, are carried out in the 
Landau gauge.

What the latest large-volume lattice studies reveal is crystal clear:
The gluon propagator of pure  Yang-Mills is infrared finite, 
both in $SU(2)$~\cite{Cucchieri:2007md} and $SU(3)$~\cite{Bogolubsky:2007ud}.
Interestingly enough, these recent lattice findings, striking as they may be, do not
constitute the first indication of this very 
characteristic behavior; several earlier simulations had found qualitatively similar 
results, even in gauges other than the Landau ( see, e.g., \cite{Alexandrou:2000ja}). 

The aforementioned lattice results, in addition to 
whatever modifications they may induce 
to other formalisms aspiring to describe the 
infrared sector of QCD, they present a serious challenge even to the practitioners 
of the gluon mass generation (albeit, a pleasant one). Indeed, 
the SDE analysis must be further refined, and freed of  
whatever major or minor theoretical shortcomings one has 
been accustomed to live with in the past.
The purpose of this talk is to report recent progress in this direction, 
and in particular on the solution of the annoying 
problem of seagull divergences~\cite{Aguilar:2009ke}, which has afflicted 
this approach from the first day of its invention~\cite{Cornwall:1982zr}.  Turns out that 
the elimination of the seagull divergences brings about an additional 
advantage, namely the separation of the SDE into two 
coupled equations,  furnishing uniquely the running of 
the effective charge and the gluon mass in the entire range of Euclidean momenta.  

\section{\label{mgsd} Seagull divergences: a perennial nuisance to gluon mass generation}
%%%%%%%%%%%%%%%%%%%%%%%%%%%%%%%%%%%%%%%%%%%%%%%%%%%%%%%%%%%%%%%%%%%%

In order to obtain  massive solutions {\it gauge-invariantly}, 
it  is necessary to invoke the well-known Schwinger mechanism~\cite{Schwinger:1962tn}.  
The basic observation is that if, for some reason, ${\bm \Pi}(q^2)$ 
acquires a pole at zero momentum transfer, then the 
 vector meson becomes massive, even if the gauge symmetry 
forbids a mass at the level of the fundamental Lagrangian.
Indeed, it is clear that if the vacuum polarization ${\bm \Pi}(q^2)$ has a pole at  $q^2=0$ with positive 
residue $\mu^2$, i.e. ${\bm \Pi}(q^2) = \mu^2/q^2$, then (in Euclidean space)
$\Delta^{-1}(q^2) = q^2 + \mu^2$.
Thus, the vector meson 
becomes massive, $\Delta^{-1}(0) = \mu^2$, 
even though it is massless in the absence of interactions ($g=0$). 
There is {\it no} physical principle which would preclude ${\bm \Pi}(q^2)$ from 
acquiring such a pole.
In a {\it strongly-coupled} theory like QCD 
this may happen for purely dynamical reasons,
since strong binding may generate zero-mass bound-state excitations~\cite{Jackiw:1973tr}.
The latter  act  {\it  like}
dynamical Nambu-Goldstone bosons, in the sense that they are massless,
composite,  and {\it longitudinally   coupled};  but, at  the same  time, they
differ  from  Nambu-Goldstone  bosons   as  far  as  their  origin  is
concerned: they  do {\it not} originate from  the spontaneous breaking
of  any global symmetry~\cite{Cornwall:1982zr}.

%%%%%%%%%%%%%%%%%%%%%%%%%%%%%%%%%%%%%%%%%%%%%%%%%%%%%%%%%%%%%%%%%%%
Of course, in order to obtain the full dynamics, 
such as, for example, the momentum-dependence of the dynamical mass, 
one must turn eventually  to the SDE that governs 
the corresponding gauge-boson self-energy. 
The way the Schwinger mechanism
is integrated into the SDE 
is through the form of the three-gluon vertex.
The latter, even in the absence of mass generation, constitutes a  
central ingredient of the SDE, and plays a crucial role in 
enforcing the transversality of the gluon self-energy.    
Therefore, an important requirement for any self-consistent Ansatz used for that 
vertex is that it should satisfy the correct  Ward identity (WI) of the PT-BFM formulation, namely  
\be
q^{\mu}{\g}_{\mu\alpha\beta}= \Delta^{-1}_{\alpha\beta}(k+q) -\Delta^{-1}_{\alpha\beta}(k)\,.
\label{VWIf}
\ee
In addition, in order to generate a dynamical mass  
one must assume that the vertex contains {\it dynamical poles}, which 
will trigger the Schwinger mechanism when 
inserted into the SDE for the gluon self-energy.

The point is that the full realization of the procedure outlined above  
is very subtle. 
In particular, even though the use of a three-gluon vertex 
containing massless poles and satisfying the correct WI
leads indeed 
to a transverse and infrared finite self-energy (i.e. $\Delta^{-1}(0) \neq 0$), 
as expected, 
the actual value of $\Delta^{-1}(0)$ has always been 
expressed in terms of divergent integrals, of the form 
(see, e.g.,\cite{Cornwall:1982zr,Aguilar:2006gr,Aguilar:2008xm})
\be
\Delta^{-1}(0) =  c_1 \int_k \Delta(k) + c_2 \int_k k^2 \Delta^2(k) \,,
\label{dqd}
\ee
where, in dimensional regularization (DR),      
\mbox{$\int_{k}\equiv\mu^{2\varepsilon}(2\pi)^{-d}\int\!d^d k$}, 
with $d=4-\epsilon$ the dimension of space-time.
This is not a problem, in principle, provided that the 
divergent integrals appearing on the rhs of (\ref{dqd})
can be properly regulated and made finite, {\it without} 
introducing counterterms of the 
form $ m^2_0 (\Lambda^2_{\chic{\mathrm{UV}}}) A^2_{\mu}$, 
which are forbidden by the local gauge invariance 
of the  fundamental  QCD Lagrangian.
Various regularization procedures of increasing sophistication 
have been tried out 
over the years, but the resulting (regularized) $\Delta^{-1}(0)$ 
remained theoretically ambiguous. Given how nicely all other pieces of the 
puzzle fit together, the underlying impression has always been 
that the ``seagull problem'' had to do  
with some (not fully understood) subtlety rather than 
some intrinsic ``need'' of the theory to produce quadratic divergences.

\section{\label{sqed} The ``seagull identity'': how to keep the photon massless (if you must)}

It is most instructive to understand what happens in a theory 
where the seagull terms do not appear due to 
the self-interactions of a gauge boson that has acquired a mass dynamically, 
but rather because the theory has   
scalar particles that are massive at tree-level. 
These scalars interact with the gauge boson, and contribute 
seagull terms to its vacuum polarization. The question is: 
if the gauge boson must remain massless, how do the 
seagull contributions disappear from that vacuum polarization?
To see how this happens, let us turn to scalar QED, 
where the aforementioned circumstances (massive scalars, must-be massless photon) are realized, 
and study the SDE governing the photon self-energy.

\begin{figure}[!t]
\begin{center}
\includegraphics[scale=0.7]{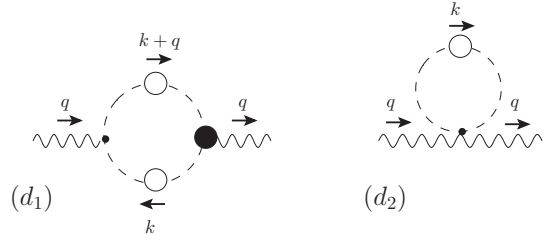}
\end{center}
\vspace{-1.0cm}
\caption{The ``one-loop dressed'' SDE for the photon self-energy.}
\label{p2}
\end{figure}

At the ``one-loop dressed'' level the SDE for the photon self-energy reads (Fig.~\ref{p2}) 
\be
\Pi_{\mu\nu}(q)= e^2 \int_k \Gamma_{\mu}^{(0)}{\mathcal D}(k){\mathcal D}(k+q)\gb_{\nu}
  + e^2 \int_k \Gamma_{\mu\nu}^{(0)}{\mathcal D}(k) \,,
\label{sself}
\ee
where ${\mathcal D}(k)$  is the fully-dressed propagator of the scalar field.
$\gb_{\nu}$ is the fully dressed photon-scalar vertex, whose 
tree-level expression is given by $\Gamma_{\mu}^{(0)}= -i(2k+q)_{\mu}$.  
Moreover, the bare quadrilinear photon-scalar vertex 
is given by $\Gamma_{\mu\nu}^{(0)}= 2ig_{\mu\nu}$. 
The  photon-scalar vertex $\Gamma_{\mu}$ and the scalar propagator  ${\mathcal D}$ 
are related by the Abelian all-order WI 
\be
q^{\nu}\gb_{\nu}= {\mathcal D}^{-1}(k+q) -{\mathcal D}^{-1}(k) \,.
\label{sward}
\ee
It is fairly easy  to demonstrate
that, by virtue of (\ref{sward}), $q^{\nu} \Pi_{\mu\nu}(q)=0$, and that $\Pi(q^2)$ reads
\be
\Pi(q^2) = \frac{-2ie^2}{d-1}
\left[ \int_k {\mathcal D}(k){\mathcal D}(k+q) k^{\mu} \gb_{\mu} - d  \int_k  {\mathcal D}(k) \right]\,.
\label{scalpi}
\ee

Let us compute from (\ref{scalpi}) the one-loop expression for $\Pi(q^2)$, to be denoted by $\Pi^{(1)}(q^2)$.
\be
\Pi^{(1)}(q^2) = \frac{-ie^2}{d-1}
\left[ \int_k (4k^2- q^2){\mathcal D}_0(k){\mathcal D}_0(k+q)  - 2d  \int_k  {\mathcal D}_0(k) \right]\,,
\label{scalpi1}
\ee
where ${\mathcal D}_0(k) = (k^2-m^2)^{-1}$. Taking the limit $q\to 0$, we find  
\be
\Pi^{(1)}(0) = \frac{-4ie^2}{d-1}
\left[ \int_k k^2 {\mathcal D}_0^2(k)  - \frac{d}{2}  \int_k  {\mathcal D}_0(k) \right]\,.
\label{scalpi10}
\ee 
Of course, there is no doubt that the 
photon remains massless perturbatively, i.e. we must have that $\Pi^{(1)}(0)=0$. 
However, the way this requirement is realized is rather subtle:
the rhs of (\ref{scalpi10}) vanishes indeed, by virtue of an identity 
that is exact in DR, namely 
\be
\int_k \frac{k^2}{(k^2-m^2)^2} = \frac{d}{2} \int_k \frac{1}{k^2-m^2}\,.
\label{id1}
\ee
Thus, the perturbative masslessness of the photon is 
explicitly realized and self-consistently enforced within the DR.
Eq.(\ref{id1}) may be cast 
in a form that is particularly suggestive for the analysis that follows, 
namely 
\be
\int_k \,k^2\frac{\partial {\mathcal D}_0(k) }{\partial k^2}= 
- \frac{d}{2} \int_k {\mathcal D}_0(k)\,.
\label{basid0}
\ee

We now return to the general Eq.(\ref{scalpi}). In order to analyze it further 
we must furnish some information about the form of $\gb_{\mu}$. Of course, any meaningful Ansatz for 
$\gb_{\mu}$ must satisfy the WI of (\ref{sward}), or else the transversality of 
$\Pi_{\mu\nu}(q)$ will be compromised from the outset. 
The form obtained by Ball and Chiu \cite{Ball:1980ay}, 
after ``solving'' the WI, under the additional  
requirement of not introducing kinematic singularities, is (we omit the 
identically conserved part of the vertex)   
\be
\gb_{\mu}= \frac{(2k+q)_{\mu}}{(k+q)^2-k^2}\left[{\mathcal D}^{-1}(k+q) -{\mathcal D}^{-1}(k)\right]\,.
\label{strans_vert}
\ee
This vertex, when substituted into (\ref{scalpi}), yields 
\be
\Pi(q^2) = \frac{ie^2}{d-1}
\left[ \int_k (4k^2- q^2)\frac{{\mathcal D}(k+q)-{\mathcal D}(k)}{(k+q)^2-k^2} +  
2d  \int_k  {\mathcal D}(k) \right]\,.
\label{scalnp}
\ee
Taking the limit of Eq.(\ref{scalnp}) as \mbox{$q\to 0$}, using that 
\be
\frac{{\mathcal D}(k+q)-{\mathcal D}(k)}{(k+q)^2-k^2} \to \frac{\partial {\mathcal D}(k)}{\partial k^2} + 
{\cal O}(q^2) \,,
\ee
we have that  
\be
\Pi(0)= \frac{4ie^2}{d-1}\left[
\int_k\, k^2\frac{\partial {\mathcal D}(k)}{\partial k^2} + \frac{d}{2}\,\int_k {\mathcal D}(k)\right].
\label{sself1}
\ee
Of course, we must have that $\Pi(0)=0$, given that there is nothing 
in the dynamics that could possibly endow the photon with a mass; in particular, the 
Schwinger's mechanism is ``switched off'', i.e. we have not introduced dynamical poles, 
and, given the form of (\ref{strans_vert}), 
neither kinematic ones, which might simulate the dynamical ones at the level of the SDE.
Thus, the rhs of (\ref{sself1}) {\it must} vanish, and therefore, we must have that 
\be
\int_k \,k^2 \frac{\partial {\mathcal D}(k)}{\partial k^2}= - \frac{d}{2}\int_k\, {\mathcal D}(k) \,,
\label{basid}
\ee
which is the non-perturbative generalization of (\ref{basid0}).

Note a crucial point: the seagull terms appearing in (\ref{sself1})
{\it cannot} be set to zero individually, 
because the scalar propagator inside them is massive (already at tree-level):
the only way to keep the photon massless, is to employ (\ref{basid}), 
which cancels them against each other. For example, if the 
term $\int_k {\mathcal D}(k)$ on the rhs were multiplied by any factor other than 
$(d/2)$ one would be stuck with seagull divergences.

\section{\label{fgmg} Massive gluons: Schwinger mechanism and seagull identity in a delicate balance}

Turns out that the construction presented in the previous section generalizes 
in the context of pure Yang-Mills theory, such as quarkless QCD, 
but {\it only} within the PT-BFM formalism!
As has been explained in detail in the recent literature~\cite{Aguilar:2006gr,Binosi:2007pi}, 
this latter formalism  
allows for a  gauge-invariant truncation of the SD series, 
in the sense that 
it preserves manifestly and at every step the transversality of the gluon self-energy.
%%%%%%%%%%%%%%%Fig.1%%%%%%%%
\begin{figure}[!t]
\begin{center}
\includegraphics[scale=0.75]{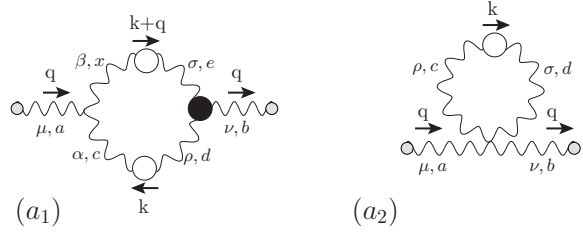}
\end{center}
\caption{``One-loop dressed'' gluonic graphs of the SDE for the PT-BFM gluon self-energy.}
\label{Fig:groupa}
\end{figure}
%%%%%%%%%%%%%%%%%%%%%%%%%%
%
Specifically, for the case at hand,  
we will consider only the ``one-loop dressed'' part of the gluon SDE that contains gluons,  
shown in  Fig.~\ref{Fig:groupa}, 
leaving out (gauge-invariantly!) the ``one-loop dressed'' ghost contributions and 
all ``two-loop dressed'' diagrams. 
Note that the Feynman rules used to build this SD series are those of the 
BFM~\cite{Abbott:1980hw};  in particular, the external gluons (distinguished by the grey circles 
attached to them) are treated as if they were background gluons. 
As we will see in a moment, 
the form of these vertices is crucial 
for obtaining from the SDE precisely the right combination of terms 
(and with the correct relative weights) that appears in  (\ref{basid}).

In order to reduce the algebraic complexity 
of the problem, we drop the longitudinal terms  
from the gluon propagators inside the integrals, i.e. we set
${\Delta}_{\alpha\beta} \to -ig_{\alpha\beta} {\Delta}$.
This does not compromise the transversality of $\widehat \Pi_{\mu\nu}(q)$ 
provided that we do the same on the rhs of the WI satisfied by ${\g}_{\nu\alpha\beta}$, namely we have simply
\be
q^{\nu} {\g}_{\nu\alpha\beta} = [\Delta^{-1}(k+q) -  \Delta^{-1}(k)]g_{\alpha\beta} \,,
\label{VWI}
\ee
instead of the full WI given in (\ref{VWIf}). 
Then, the SDE corresponding to Fig.~\ref{Fig:groupa} reduces to 
\be
\widehat{\Delta}^{-1}(q)  =
q^2  + \frac{i g^2 C_{\rm {A}}}{2(d-1)} 
\bigg[ \int_k \widetilde{\Gamma}_{\mu\alpha\beta}^{(0)}
\Delta (k) \Delta(k+q) {\g}^{\mu\alpha\beta}
+  2 d^2 \int_k \!\Delta(k)\bigg],
\label{SDEm}
\ee
where $C_{\rm {A}}$ the Casimir eigenvalue 
of the adjoint representation [$C_{\rm {A}}=N$ for $SU(N)$]. The vertex 
\mbox{$\widetilde{\Gamma}_{\mu\alpha\beta}^{(0)} (q,k,-k-q)$} 
is the bare three-gluon vertex in the Feynman gauge of the BFM,  
and ${\g}_{\mu\alpha\beta}$ denotes its fully-dressed version. 

The function $\widehat{\Delta}(q)$ appearing on the lhs of (\ref{SDEm})
is the scalar part of the gluon propagator 
in the BFM, i.e. two background gluons entering. $\widehat{\Delta}(q)$   
is related to the 
standard $\Delta(q)$, defined in the $R_{\xi}$ gauges, by 
means of the powerful identity  $\widehat{\Delta}(q) [1+G(q^2)]^2 = \Delta(q)$, where  
$G(q^2)$ is an auxiliary two-point function~\cite{Grassi:1999tp}, 
which, quite remarkably, coincides in the Landau gauge with the well-known 
Kugo-Ojima function (see, e.g. talk of Daniele Binosi in these proceedings~\cite{Binosi:2009er}).
We will next set $G(q^2)=0$, i.e. 
we effectively assume that, inside the quantum loops, $\Delta(q)=\widehat{\Delta}(q)$.

At this point enters the new ingredient: 
a judicious Ansatz for the three-gluon vertex
which, in addition to satisfying (\ref{VWI}) will 
allow us to use the seagull identity (\ref{basid}) and get a non-vanishing and finite  $\Delta^{-1} (0)$. 
 
To begin with, let us first write $\Delta^{-1} (q)$ in the alternative form (in Minkowski space)
\be
\Delta^{-1} (q) = q^2 {H}^{-1}(q) - {\widetilde m}^2(q) \,.
\label{ombe}
\ee
The tree-level result for $\Delta^{-1} (q)$ is recovered by setting ${H}^{-1}(q)=1$ and ${\widetilde m}^2=0$. 
Then, an appropriate Ansatz for ${\g}_{\nu\alpha\beta}$ is given by~\cite{Aguilar:2009ke}   
\be
i{\g}_{\mu\alpha\beta} = 
\left[\frac{(k+q)^2 {H}^{-1}(k+q) - k^2 {H}^{-1}(k)}
{(k+q)^2-k^2}\right]\widetilde{\Gamma}_{\mu\alpha\beta}^{(0)} \,+ \,V_{\mu\alpha\beta} \,,
\label{full3}
\ee
where the term $V_{\mu\alpha\beta}$ contains the non-perturbative contributions 
due to bound-state poles associated with the Schwinger mechanism. 
Note that we must have   
\be
q^{\mu} V_{\mu\alpha\beta} = [{\widetilde m}^2(k) - {\widetilde m}^2(k+q)] g_{\alpha\beta}\,,
\label{mvertwi}
\ee
in order for the ${\g}_{\mu\alpha\beta}$ of Eq.~(\ref{full3}) to satisfy (by construction) the correct WI of (\ref{VWI}).

\begin{figure}
\bce
\includegraphics[scale=0.5]{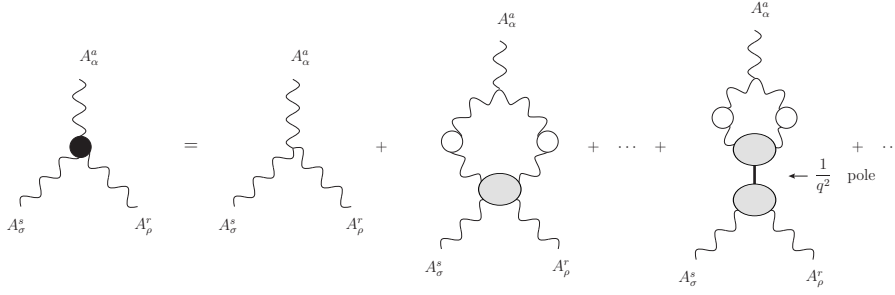}
\ece
\vspace{-0.5cm}
\caption{The SDE for the three-gluon vertex. All kernels are one-particle irreducible, and the  
$1/q^2$ pole is not kinematic but dynamical (purely non-perturbative);
physically it corresponds to a (composite) Goldstone mode, necessary for maintaining the local gauge invariance.}
\label{jsv}
\end{figure}
The Ansatz of (\ref{full3}) 
mimics that of Eq.~(\ref{strans_vert})  
to the extent that the first term contains the right structure 
to produce, when inserted into the first term on the rhs of (\ref{SDEm}), the  derivative term 
appearing on the lhs of (\ref{basid}). The rhs of (\ref{basid})
is already there: it is the second term on the rhs of (\ref{SDEm}), 
originating directly from the seagull diagram ($a_2$). 

Similarly, a simple Ansatz for $V_{\mu\alpha\beta}$
that captures the two essential characteristics 
of having a (composite), longitudinally coupled poles, 
and satisfying the WI of  (\ref{mvertwi}) is 
\be
V_{\mu\alpha\beta} = V_{\mu\alpha\beta}^{\rm{\ell}} + V_{\mu\alpha\beta}^{\rm{t}}\,,
\label{mvert}
\ee
where
\be
V_{\mu\alpha\beta}^{\rm{\ell}} = \frac{q_{\mu}}{q^2} 
\bigg[{\widetilde m}^2(k) - {\widetilde m}^2(k+q)\bigg] g_{\alpha\beta} \,,
\label{mvert2}
\ee
and with the transverse part $V_{\mu\alpha\beta}^{\rm{t}}$ 
satisfying  $q^{\mu}V_{\mu\alpha\beta}^{\rm{t}} =0$.
We can write the vertex of (\ref{full3}) 
equivalently as 
\be
i{\g}_{\mu\alpha\beta} = 
\left[ \frac{\Delta^{-1}(k+q) -  
\Delta^{-1}(k)}{(k+q)^2-k^2} \right]\widetilde{\Gamma}_{\nu\alpha\beta}^{(0)}
\, + \,{\overline V}_{\mu\alpha\beta} \,,
\label{full4}
\ee
with 
\be
{\overline V}_{\mu\alpha\beta} = V_{\mu\alpha\beta} + V_{\mu\alpha\beta}^{\rm{r}}\,,
\ee
where 
\be
V_{\mu\alpha\beta}^{\rm{r}} = 
(2k+q)_{\mu} \left[ \frac{{\widetilde m}^2(k+q) -{\widetilde m}^2(k) }{(k+q)^2-k^2} \right]g_{\alpha\beta}\,.
\label{efvsk}
\ee
The term $V_{\mu\alpha\beta}^{\rm{r}}$ is a residual piece, acting as an additional (non-perturbative) vertex term, 
originating from forcing the vertex to assume the form of (\ref{full4}). 
This last way of writing ${\g}^{\nu\alpha\beta}$ 
makes the use of the basic identity of Eq.(\ref{basid}) immediate.
Thus, after these rearrangements, we have that the final non-perturbative 
effective vertex ${\overline V}_{\mu\alpha\beta}$ must be transverse,  
$q^{\mu} {\overline V}_{\mu\alpha\beta}  = 0.$

Substituting 
for the ${\g}^{\mu\alpha\beta}$ on the rhs the expression
given in  (\ref{full4}) we obtain after simple algebra 
\be
{\Delta}^{-1}(q^2)  =
q^2  - \frac{i g^2 C_{\rm {A}}}{2(d-1)} \bigg[\Pi(q)+ \Pi_{\widetilde m}(q)\bigg] \,,
\label{hmt}
\ee
with 
\be
\Pi(q) =    
(7d-8)\, q^2 \int_k \frac{\Delta(k+q)- \Delta(k)}{(k+q)^2-k^2} 
+4d \bigg[\int_k k^2 \, \frac{\Delta(k+q)- \Delta(k)}{(k+q)^2-k^2}+ \frac{d}{2} \int_k \!\Delta(k) \bigg] \,,
\label{Piq}
\ee
and
\be
\Pi_{\widetilde m}(q) = \int_k \widetilde{\Gamma}^{(0)}_{\mu\alpha\beta}\Delta (k) \Delta(k+q)
[V^{\rm{\ell}} + \{V^{\rm{t}} + V^{\rm{r}}\} ]^{\mu\alpha\beta}\,.
\label{Pim}
\ee
The term in square brackets on the rhs of (\ref{Piq}) has exactly the structure needed for employing (\ref{basid}).
Note the perfect balance of relative coefficients required for this to happen!
This becomes possible within the PT-BFM framework thanks to the special vertices appearing in the SDE; 
instead, in the conventional SD formulation (e.g., in the $R_{\xi}$ gauges)
it would be very difficult to obtain 
the precise combination of terms needed for implementing (\ref{basid}). 
Evidently, by virtue of (\ref{basid}) 
it is clear that $\Pi(0)=0$. Thus, the part of the calculation 
determining $\Pi(q)$ is very similar to that of scalar QED, in the sense that 
it leads to {\it total seagull annihilation}, keeping the gluon massless. 

On the other hand, the term $\Pi_{\widetilde m}(q)$, not present in the scalar QED study, 
makes it possible to have ${\Delta}^{-1}(0) \neq 0$ for the gluons. 
Assuming, for simplicity, that the dominant contribution in (\ref{Pim}) comes from $V^{\rm{\ell}}$,  
we obtain 
\be
\Pi_{\widetilde m}(q) =
-\frac{2 d }{q^2}\int_k {\widetilde m}^2(k) \Delta(k)\Delta(k+q)[(k+q)^2-k^2]\,.
\label{pim}
\ee
Now, in the limit $q^2 \to 0$ (in Euclidean space)  we have that 
\be
\lim_{q^2\to0} \Bigg\{\frac{1}{q^2}\int_{k_{\chic E}} {\widetilde m}^2(k) \Delta(k)\Delta(k+q)[(k+q)^2-k^2]\Bigg\} = 
- \frac{1}{2}\int_{k_{\chic E}} k^2 \Delta^2(k^2) [{\widetilde m}^{2}(k^2)]^{\prime} \,.
\label{pimc} 
\ee
where the ``prime'' denotes differentiation with respect to $k^2$. 
Note that a monotonically decreasing mass,  
$ [{\widetilde m}^{2}(k^2)]^{\prime}<0$,  guarantees 
the positivity of ${\widetilde m}^{2}(0)$ (in Euclidean space)

An important consequence of this analysis is that Eq.~(\ref{hmt}) 
can be split unambiguously into two parts, one that vanishes as 
$q^2 \to 0$ and one that does not. In fact, using (\ref{ombe}) on the lhs of (\ref{hmt}), 
we can assign the two types of contributions into two separate (but coupled) equations, 
namely (Minkowski space)  
\bea
q^2 {H}^{-1}(q) &=& q^2 - \frac{i g^2 C_{\rm {A}}}{2(d-1)} \Pi(q)\,,
\label{sep1}\\
{\widetilde m}^2(q) &=& \frac{i g^2 C_{\rm {A}}}{2(d-1)} \Pi_{\widetilde m}(q)\,.
\label{sep2}
\eea
As we will see shortly, the first equation will determine the momentum dependence of the 
effective charge, and the second the running of the gluon mass. 

\section{\label{coupeq} Effective charge and gluon mass: coupled but unique}

It is well-known that, due to the Abelian WIs of the PT-BFM Green's functions, the product 
\be
{\widehat d}_0(q^2) = g^2_0 \widehat\Delta_0(q^2) = g^2 \widehat\Delta(q^2) = {\widehat d}(q^2), 
\label{ptrgi}
\ee
forms a renormalization-group (RG)-invariant \mbox{($\mu$-independent)} quantity~\cite{Cornwall:1982zr}.
In order to realize Eq.(\ref{ptrgi}) non-perturbatively, first set  
\be
{\widetilde m}^2(q^2) = m^2(q^2) H^{-1}(q^2) \,,
\label{rg2}
\ee
where $m^2(q^2)$ is assumed to be the RG-invariant dynamical gluon mass. 
Then 
\be
{\widehat\Delta}(q^2) = \frac{H(q^2)}{q^2 + m^2(q^2)}\,,
\label{rg3}
\ee
and from the requirement that  
$g^2 {\widehat\Delta}(q)$ must be RG-invariant we have that 
\be
g^2 H(q^2) = {\overline g}^2(q^2)\,.
\label{rg1}
\ee
Therefore, we finally arrive at the RG-invariant combination 
\be
{\widehat d}(q^2)\equiv g^2{\widehat\Delta}(q^2) = {\overline g}^2(q^2)\bar\Delta(q^2)\,,
\label{rg4}
\ee
with
\be
\bar\Delta(q^2)= \frac{1}{q^2 + m^2(q^2)}\,.
\label{rg5}
\ee
So, ${\widehat d}(q^2)$ is written as the product of two RG-invariant quantities: 
the dimensionless running coupling ${\overline g}^2(q^2)$ and the dimensionful 
``massive'' gluon propagator $\bar\Delta(q^2)$.

We next cast our analysis in terms of the RG-invariant quantities defined above.
The use of the spectral representation~\cite{Cornwall:1982zr,Cornwall:2009ud} for $\Delta(q^2)$, namely 
\begin{equation}
\Delta (q^2) = \int \!\! d \lambda^2 \, \frac{\rho\, (\lambda^2)}{q^2 - \lambda^2 + i\epsilon}\,, 
\label{lehmann}
\end{equation} 
results in a spectacular simplification, because it ``solves'' the combination $\{\frac{\Delta(k+q)- \Delta(k)}{(k+q)^2-k^2}\}$ 
appearing in (\ref{Piq}). After a series of standard assumptions   
one finally obtains~\cite{Aguilar:2009ke} 
\be
\frac{1}{{\overline g}^2(q^2)} =  \frac{1}{{\overline g}^2(\mu^2)} 
+ \tilde{b}\left[\int^{q^2\!/\!4}_{0}\!\!\!dz \left(1+ \frac{4z}{5q^2}\right) \left(1-\frac{4z}{q^2}\right)^{\!\! 1/2}\!\!\!\! \bar\Delta(z)  
- \int^{\mu^2\!/\!4}_{0}\!\!\!dz  \left(1+ \frac{4z}{5\mu^2}\right)\left(1-\frac{4z}{\mu^2}\right)^{\!\! 1/2}\!\!\!\! \bar\Delta(z)\right]\,,
\label{eqalphargi}
\ee
and
\be
\frac{m^2(q^2)}{{\overline g}^2(q^2)} = \frac{2 \tilde{b}}{5} 
\bigg[
\bar\Delta(q^2) \int_0^{q^2} \!\!\! dy y m^2(y) \bar\Delta(y) 
\,-\, \frac{1}{2} 
\int_{q^2}^{\infty}\!\!\! dy y^2 {\bar\Delta}^2(y) {\overline g}^2(y) [m^{2}(y)]^{\prime} 
\bigg]\,,
\label{ppgtm}
\ee
where $\tilde{b} = 10 C_{\rm {A}}/ 48\pi^2 $; the discrepancy from the  
factor $b = 11 C_{\rm {A}}/ 48\pi^2 $, namely the first coefficient 
of the QCD one-loop $\beta$-function, is due to the (gauge-invariant!) omission 
of the ghost loops.

\begin{figure}[!t]
\begin{center}
\includegraphics[scale=2.0]{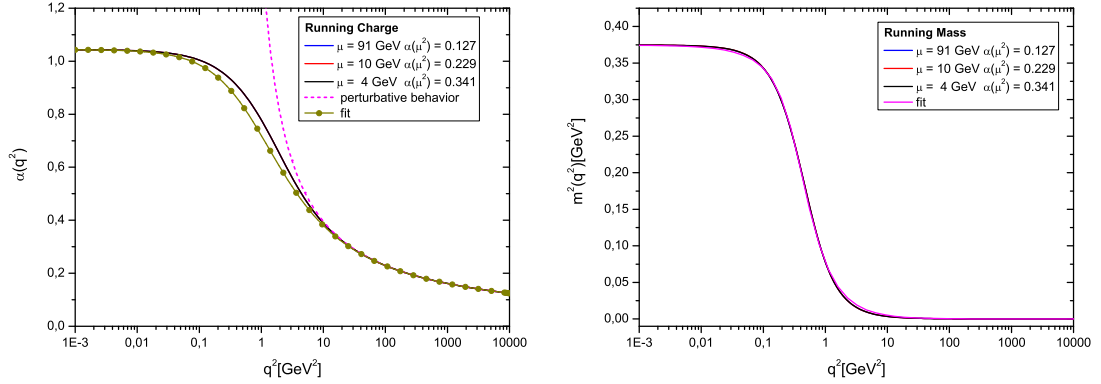}
\end{center}
\vspace{-1.0cm}
\caption{Numerical solutions displaying the momentum-dependence  of the effective charge and the gluon mass, 
for three different values of the renormalization point $\mu$.}
\label{plot1}
\end{figure}

To study the behavior of the solutions of (\ref{ppgtm}) for asymptotically large $q^2$, 
set ${\bar\Delta}(x)\to 1/x$ and ${\bar\Delta}(y)\to 1/y$ to arrive at 
\be
m^2(q^2) \ln q^2 =  \frac{2}{5} 
\bigg[ \frac{1}{q^2} \int_0^{q^2} dy \, m^2(y) \, - \,\frac{1}{2} 
\int_{q^2}^{\infty} \!\!\! dy {\overline g}^2(y) [m^{2}(y)]^{\prime}
\bigg]\,.
\label{pls}
\ee
It is relatively straightforward to establish that the asymptotic solutions of (\ref{pls}) display power-law running.
Indeed, substituting on both sides of (\ref{pls}) a $m^2(q^2)$ of the form  
\be
m^2(q^2)  = \frac{\lambda_0^4}{q^2} (\ln q^2)^{\gamma-1}\,, 
\label{plr}
\ee
one recognizes  
that the second term on the rhs of (\ref{plr}) is subleading, 
and that (\ref{plr}) is a solution of (\ref{pls}) provided that $\gamma = \frac{2}{5}$. 

%%%%%%%%%%%%%%%%%%%%%%%%%%%%%%%%%%%%%%%%%%%%%%%
%%%%%%%%%%%%%%%%%%%%%%%%%%%

\begin{figure}[!t]
\begin{center}
\includegraphics[scale=0.9]{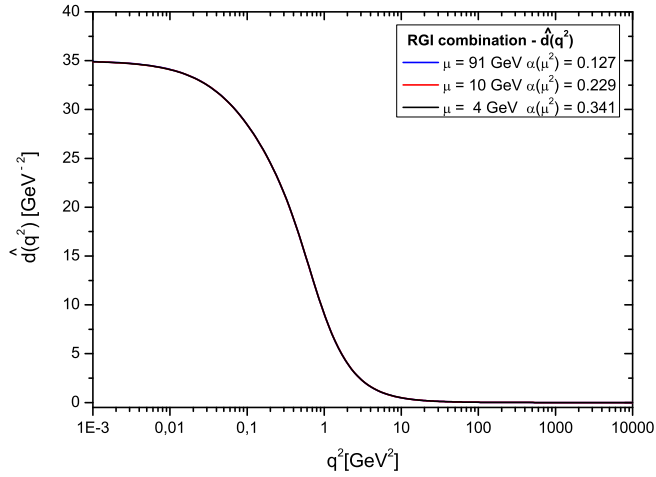}
\end{center}
\vspace{-1.0cm}
\caption{The RG-invariant product $\widehat{d}(q^2)$ obtained by combining the 
results for $\alpha(q^2)$ and $m^2(q^2)$.} 
%according to Eq.~(\ref{rg4}).}
\label{plot3}
\end{figure}

We next solve numerically the two coupled integral equations, 
renormalizing them at three different
points, namely $\mu=\{4,10,91\}\,\mbox{GeV}$, with $\alpha(\mu^2)= g^2(\mu)/4\pi =\{0.341,0.229,0.127\}$, respectively. 
In Fig.~\ref{plot1}, we show the results for $\alpha(q^2)$
and $m^2(q^2)$;  for either quantity we see clearly that 
the three curves merge practically into a single one, thus 
confirming numerically their $\mu$-independence, expected on formal grounds. 
The solutions for $\alpha(q^2)$ may be fitted  by the 
physically motivated functional form~\cite{Cornwall:1982zr}, namely   
\be
\alpha(q^2) = \frac{1}{4\pi\tilde{b}\ln[(q^2+tm_0^2)/\Lambda^2]} \,,
\label{fit_coup}
\ee
with $t=3.7$ and $\Lambda= 645\,\mbox{MeV}$. 
The behavior of $m^2(q^2)$ in the entire range of momenta can be accurately described by the following parametrization
\be
m^2(q^2)=\frac{m_0^4}{q^2+m_0^2}\left[\ln\left(\frac{q^2+f(q^2,m_0^2)}{\Lambda^2}\right)\bigg/
\ln\left(\frac{f(0,m_0^2)}{\Lambda^2}\right)\right]^{-3/5} \,,
\label{fit_mass}
\ee
where 
$f(q^2,m_0^2)= \rho_1 m_0^2 + \rho_2 \frac{m_0^4}{q^2 + m_0^2}$,
with $\rho_1=-1/2$, $\rho_2=5/2$, and $m_0=612\,\mbox{MeV}$.
Finally, with the help of Eq.~(\ref{rg4}) we can construct $\widehat{d}(q^2)$ 
out of the numerical solutions for $\alpha(q^2)$ and $m^2(q^2)$; the result is 
shown in Fig.~{\ref{plot3}}. Obviously, since $\widehat{d}(q^2)$ is built out of two quantities that are individually independent of $\mu$, 
it too turns out to be  $\mu$-independent; 
this property is clearly observed in  Fig.~\ref{plot3}.

\section{Conclusions}

The analysis presented here demonstrates that 
the appearance of seagull divergences in gluon mass generation is 
caused by a subtle  
mismatch between two parallel field theoretic mechanisms.  
Specifically, the Schwinger 
mechanism, which requires the appearance of massless poles in the three-gluon vertex, 
distorts the mechanism responsible for the 
cancellation of the seagull divergences, {\it unless} the poles enter into the gluon vertex 
in a very particular way. A concrete example of a vertex that does not produce any 
clash between these two mechanism has been given, and the implications for 
the resulting SDE have been worked out.
The elimination of the seagull divergences allows the resulting SDE equation to be 
separated unambiguously into two distinct dynamical equations, determining the 
gluon mass and the QCD effective charge. This, in turn, is a significant improvement 
over the existing approaches, where the infrared behavior of 
these two quantities had to be extracted (not without a certain ambiguity) from $\widehat{d}(q^2)$.  
It is clear that the methodology outlined here should be applied to  
gauges where the results can be directly 
compared to lattice simulations (such as the Landau gauge). 
Given that the solutions for  $\alpha(q^2)$ and $m^2(q^2)$, 
and therefore for  $\widehat{d}(q^2)$ and/or $\Delta(q^2)$, are expected to be 
unique, one should be able to test if the freezing value  $\Delta(0)$ obtained within  
this new SDE treatment coincides with that seen on the lattice. We hope to report 
progress in this direction in the near future.

{\it Acknowledgments:} 

I would like to thank the ECT* for making the QCDTNT workshop possible. 
This research was supported by the European FEDER and  Spanish MICINN under grant FPA2008-02878,
the program Prometeo/2009/129 (Generalitat Valenciana), and the Fundaci\'on General of the UV.

\end{document}